\documentclass[twocolumn]{autart}    
\usepackage{amsmath,amssymb,amsfonts}
\usepackage{lipsum,cuted}
\usepackage{dsfont}
\usepackage{graphicx}
\usepackage{algorithm,algorithmic}
\usepackage{color}
\usepackage{cite}
\usepackage{subcaption}
\usepackage{epstopdf}
\usepackage{bm}
\usepackage{mleftright}
\usepackage{mathrsfs}
\usepackage{stmaryrd}
\usepackage{apacite}

\newtheorem{theorem}{Theorem}
\newtheorem{proposition}{Proposition}

\newtheorem{lemma}{Lemma}
\newtheorem{corollary}{Corollary}
\newtheorem{definition}{Definition}

\newtheorem{remark}{Remark}
\newtheorem{assumption}{Assumption}
\newtheorem{example}{Example}

\newcommand{\black}[1]{{\color{black}{#1}}}

\def\Re{\mbox{\rm Re}}
\def\Im{\mbox{\rm Im}}

\def\bmP{\mbox{$\bm{P}$}}
\def\bmG{\mbox{$\bm{\Gamma}$}}
\def\bmC{\mbox{$\bm{C}$}}

\def\nlcls{\mbox{$\bm{P}\,\#\,\bm{C}$}}
\def\mr+{\mbox{$\text{mr}_{+}$}}

\graphicspath{{./fig/}}
\begin{document}

\begin{frontmatter}

\title{
A Frequency-Domain Approach to Nonlinear Negative Imaginary Systems Analysis\thanksref{footnoteinfo}}

\thanks[footnoteinfo]{This work was supported in part by Shanghai Municipal Science and Technology Major Project under grant  2021SHZDZX0100. }

\author[TJ,ost]{Di Zhao}\ead{dzhao925@tongji.edu.cn},    
\author[HKUST]{Chao Chen}\ead{cchenap@connect.ust.hk},   
\author[SZ]{Sei Zhen Khong}\ead{szkhongwork@gmail.com},    
   
\thanks[ost]{Corresponding author.}
\address[TJ]{Department of Control Science and Engineering \& Shanghai Institute of Intelligent Science and Technology, Tongji University, Shanghai, China}
\address[HKUST]{Department of Electronic and Computer Engineering, The Hong Kong University of Science and Technology, Clear Water Bay, Kowloon, Hong Kong, China}  
\address[SZ]{Independent Researcher}  

\vspace{-5pt}

\begin{abstract}                          
In this study, we extend the theory of negative imaginary (NI) systems to a nonlinear framework using a frequency-domain approach. The extended notion is completely characterized via a finite-frequency integration over a ``kernel function'' on energy-bounded input and output signal pairs. The notion is closely related to and carefully contrasted with the well-studied extension of negative imaginariness --- the theory of counterclockwise dynamics. A condition for feedback stability of the proposed nonlinear NI systems is then developed based on the technique of integral quadratic constraints. Examples and simulations on feedback interconnections of typical nonlinear systems are provided to demonstrate the effectiveness. 

\end{abstract}

\end{frontmatter}

\section{Introduction}
Motivated in part by the robust vibration control of systems with flexible structures, the theory of negative imaginary \black{(NI)} systems was first proposed by \cite{Lanzon2008TAC,Petersen2010MCS}. The notion of negative imaginariness, characterizing certain input-output properties of dynamical systems, is a natural parallel to positive realness \cite{brogliato2007,brian2006Network,arjan2017L2Gain}. 
NI systems are crucial for engineering applications, especially for those of lightly damped structures with colocated position sensors and force actuators \cite{Lanzon2008TAC,Petersen2010MCS,Ian2014TAC,Ian2016AnnualRev}. 

The well-studied counterclockwise (CCW) input-output dynamics, which can be regarded as a nonlinear generalization of NI systems, were firstly studied in \cite{angeli2006CCW,angeli2007multistability}. Recently, with the blossoming of the NI systems theory (see, e.g.,
 \cite{Ian2016AnnualRev,KhongPR17,kurawa2020negative,Mabrok2021AEJ} and the references therein), there have been attempts on generalizing the theory from various perspectives, among which extending the theory to incorporating nonlinear systems is of great interest and importance. Using dissipativity theories, \cite{Petersen2018CDC,Mabrok2021AEJ} extends the framework to nonlinear systems from a state-space perspective. By introducing a notion called phases of systems, \cite{chen2021phase} extends a special version of the theory to phase-bounded (linear time-invariant) LTI systems and \cite{chao2020nonlinear} extend it to phase-bounded nonlinear operators. As a summary, there have been mainly two types of extensional studies for the NI systems theory, with one using state-space methods \cite{Petersen2018CDC,Mabrok2021AEJ} and the other from an  input-output perspective \cite{angeli2006CCW,chen2021phase,chao2020nonlinear}. All of these extensions have advantages and disadvantages in different applications and scenarios, leaving us to wonder what the most ``natural'' way is for the extension to nonlinear systems. This also becomes one of the main motivations for this study, in which we propose an extension of the NI systems theory to nonlinear systems from an input-output perspective via a frequency-domain approach. To be precise, under certain mild conditions, the proposed negative imaginariness of a nonlinear system can be verified purely by the finite-frequency properties of the input-output signal pairs of the system. {One important reason to extend the framework using frequency-domain approaches lies in that the NI systems theory was initially studied on transfer functions. Along the line of frequency-domain methods, the extended theory can be easily and naturally traced back to its original form.}
 
The elegant results on feedback interconnections of open-loop NI systems are the most appealing and crucial parts of the existing theory and its applications. The robust feedback stability of NI systems was investigated in \cite{Lanzon2008TAC} as a parallel to the positive real 
stability results \cite{brogliato2007}. \cite{Petersen2018CDC} obtained feedback stability conditions for ``nonlinear NI'' systems using state-space methods and dissipativity theories, while \cite{chao2020nonlinear} explored similar conditions by proposing a ``nonlinear small phase theorem'' using techniques on graph separation and multiplier approaches. In this study, by extending stability results in \cite{KhongPR17}, we obtain feedback stability of nonlinear NI systems by imposing suitable integral quadratic constraints (IQCs) \cite{megretski1997system,rantzer1997integral,Khong2013TAC,cantoni2013SICON,khong2021IQC} for input-output signal pairs \black{around zero or infinity frequency}. 
 
The rest of the paper is organized as follows. In Section~\ref{sec:notation}, the basic notation and preliminary results on systems are introduced. In Section~\ref{sec:NI}, we propose a definition of nonlinear NI systems and its strict version, and compare them with existing definitions of CCW dynamics. The main result --- a robust feedback stability condition for NI systems --- is obtained in Section~\ref{sec:main}. Illustrative examples and demonstrating simulations are provided in Section~\ref{sec:sim}. Finally, the study is concluded in Section~\ref{sec:conc}.

\section{Notation and Preliminaries}\label{sec:notation}
\subsection{Basic Notation}
Let $\mathbb{F} = \mathbb{R}$ or $\mathbb{C}$ be the real or complex field,  and $\mathbb{F}^n$ be the linear space of $n$-tuples of $\mathbb{F}$ over the field $\mathbb{F}$. For $x,y \in \mathbb{F}^n$, the inner product is denoted by $\langle x,y\rangle$ and the Euclidean norm by $|x|:=\sqrt{\langle x,x\rangle}$. The real and imaginary parts of a complex number $s\in\mathbb{C}$ are denoted by $\Re\,s$ and $\Im\,s$, respectively, and its conjugate by $\bar{s}$. The complex conjugate transpose of a matrix $A\in\mathbb{C}^{n\times n}$ is denoted by $A^*$, its conjugate transpose by $A^T$, and its singular values by
$$\bar{\sigma}(A)= \sigma_1(A) \geq\sigma_{2}(A)\geq \cdots \geq \sigma_{n}(A)=\underline{\sigma}(A).$$

Denote the set of all absolutely integrable signals by 
$$\mathcal{L}^n_1
:= \left\{u : [0, \infty) \to \mathbb{R}^n\Big|~ \int_0^\infty |u(t)| \, dt < \infty\right\}.$$
Denote the set of all energy-bounded signals by
$$\mathcal{L}^n_2
:= \left\{u : [0, \infty) \to \mathbb{R}^n\Big|~ \|u\|_2^2 := \int_0^\infty |u(t)|^2 \, dt < \infty\right\}.$$
For $u\in\mathcal{L}^n_2$, its Fourier transform is denoted by $\hat{u}$. For $T\geq 0$, define the truncation operator $\bm{\Gamma}_{T}$ on all signals $u : [0, \infty) \to \mathbb{R}^n$ by
$$(\bm{\Gamma}_{T} u)(t) = \left\{
\begin{array}{ll}
u(t), & \hbox{$0\leq t\leq{T}$;} \\
0, & \hbox{otherwise.}
\end{array}
\right.
$$
Denote the extended $\mathcal{L}_2$ space as
$$\mathcal{L}^n_{2e}:=\left\{u :[0, \infty) \to \mathbb{R}^n|~ \bm{\Gamma}_{T} u\in\mathcal{L}^n_2,~\forall~T>0\right\}.$$
Denote by $\mathcal{L}_\infty$ the Lebesgue $\infty$-space of functions that are essentially bounded on the imaginary axis $j\mathbb{R}$. Denote by $\mathcal{H}_\infty$
the Hardy $\infty$-space of functions that are holomorphic and uniformly bounded on the open right-half complex plane. The $\mathcal{H}_\infty$ norm of a function $\hat{G}\in\mathcal{H}_\infty^{n\times n}$ is defined as
$$\|\hat{G}\|_\infty:=\sup_{\text{\rm Re}\,s>0}\bar{\sigma}(\hat{G}(s))=\sup_{\omega\in\mathbb{R}}\bar{\sigma}(\hat{G}(j\omega)).$$
Denote by $\mathcal{RH}_\infty^{n\times n}$ the set of all real rational members in $\mathcal{H}_\infty^{n\times n}$. A linear time-invariant (LTI) system with transfer matrix $\hat{G}$ is said to be stable if $\hat{G}\in\mathcal{RH}_\infty^{n\times n}$. In what follows, the superscripts in $\mathcal{H}_\infty^{n\times n}$, $\mathcal{L}^n_{2e}$ $\dots$ will be omitted when the context is clear, and so will the frequency-domain symbol $s$ or $j\omega$. 

The following definition on negative imaginariness for stable LTI systems is taken from \cite{Lanzon2008TAC}. 
\begin{definition}[LTI Negative Imaginary Systems]\label{def:LTI_NI}
	A stable LTI system with transfer matrix $\hat{G}\in\mathcal{RH}^{n\times n}_\infty$ is said to be negative imaginary (NI) if 
	$$j\left(\hat{G}(j\omega)-\hat{G}(j\omega)^*\right) \geq 0,~\forall \omega\in(0,\infty).$$
	It is said to be strictly negative imaginary (SNI) if
	$$j\left(\hat{G}(j\omega)-\hat{G}(j\omega)^*\right) > 0,~\forall \omega\in(0,\infty).$$
\end{definition}

\subsection{Nonlinear Systems}

We regard a nonlinear system as an operator mapping from $\mathcal{L}^n_{2e}$ to $\mathcal{L}^n_{2e}$ in this study. 
\begin{definition}[Causality]
	A system $\bm{P}:~\mathcal{L}_{2e}^n\to \mathcal{L}^n_{2e}$ is said to be causal if for all ${T}> 0$ and $u_1,u_2 \in \mathcal{L}^n_{2e}$,
	$$\bm{\Gamma}_{T} u_1 = \bm{\Gamma}_{T}  u_2 \Rightarrow \bm{\Gamma}_{T} \bm{P} u_1 = \bm{\Gamma}_{T} \bm{P} u_2.$$
\end{definition}

The $\mathcal{L}_2$ domain of a causal system $\bmP$ is defined as
$$\mathcal{D}(\bmP):=\{u\in\mathcal{L}^n_2|~\bmP u\in\mathcal{L}^n_2\}.$$
Denote by $\mathcal{C}^n$ the set of all absolutely continuous functions, which are differentiable almost everywhere, in $\mathcal{L}^n_{2e}$. 
In this study, we mainly investigate nonlinear operators in the following set \begin{align*}\mathcal{N}_n:=\{\bmP:~\mathcal{L}^n_{2e}\to\mathcal{L}^n_{2e}|\bmP 0=0,~\bmP~\text{is causal}\}.\end{align*}
Denote a subset of the above operators with output signals being absolutely continuous as
$$\mathcal{N}^{\mathcal{C}}_n:=\{\bmP\in\mathcal{N}_n|~\bmP\mathcal{L}_{2e}^n\subset\mathcal{C}^n\}.$$
\begin{definition}[Stability]\label{def:finite_gain_stable}
	A system $\bm{P}\in\mathcal{N}_n$ is said to be (finite-gain) stable if there exists $\alpha>0$ such that
	\begin{align}\label{eq:def_finite_gain_stable}
	\|\bmG_T\bmP u\|_2\leq \alpha\|\bmG_T u\|_2,~\forall~T\geq 0,~u\in\mathcal{L}_{2e}^m.
	\end{align}
\end{definition}
\black{The following lemma is a direct consequence of \cite[Proposition~1.2.3]{arjan2017L2Gain}.
\begin{lemma}\label{lem:finite_gain_stability_def}
	A system $\bmP$ is finite-gain stable if and only if $\mathcal{D}(\bm{P})=\mathcal{L}^m_2$ and
	\begin{align*}
		\|\bm{P}\| := \sup_{0 \neq u\in \mathcal{L}^m_2}\frac{\|\bm{P}u\|_2}{\|u\|_2}< \infty.
	\end{align*}
\end{lemma}
}
\begin{figure}
	\centering
	\includegraphics[scale=0.58]{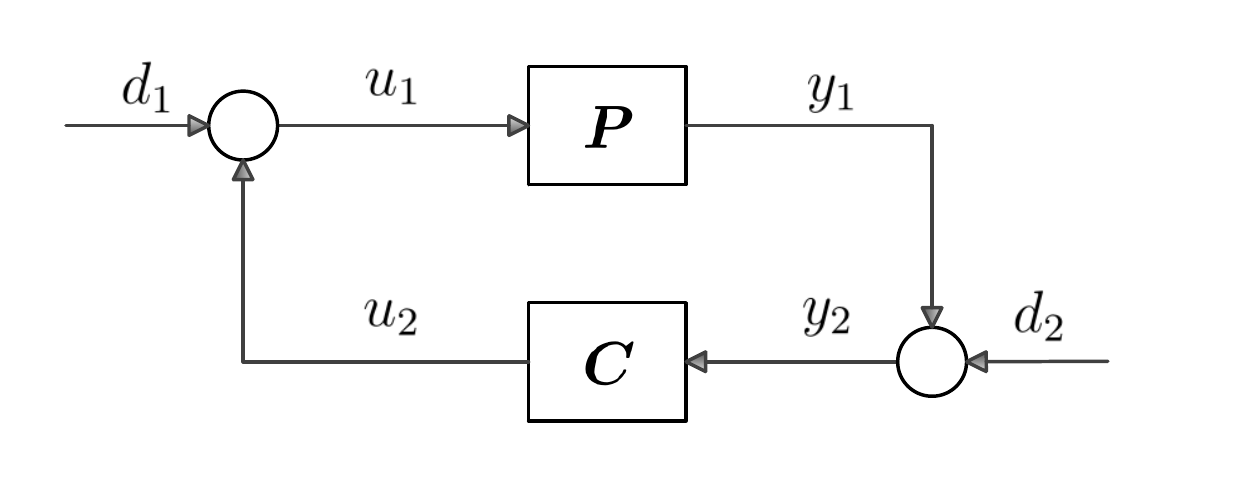}\\
	\caption{A standard closed-loop system $\nlcls$.}\label{figloop}
\end{figure}
Feedback systems and their stability are the main focus of this study. Denote by $\nlcls$ the positive feedback system, as shown in Fig.~\ref{figloop}, between $\bmP\in\mathcal{N}_n$ and $\bmC\in\mathcal{N}_n$. 

In this study, we adopt the well-posedness definition from \cite{Willems1971nonlinear,Vidyasagar1993nonlinear}.
\begin{definition}[Well-posedness]\label{def:wellposed}
	The closed-loop system $\nlcls$ in Fig.~\ref{figloop} is said to be well-posed~if
	\begin{align*}
		\bm{F}_{\bm{P},\bm{C}}:&~\mathcal{L}_{2e}^n\times \mathcal{L}_{2e}^n \to \mathcal{L}_{2e}^{n}\times \mathcal{L}_{2e}^n \\
		&=\begin{bmatrix}
			u_1 \\
			y_2 \\
		\end{bmatrix} \mapsto \begin{bmatrix}
			d_1 \\
			d_2 \\
		\end{bmatrix}= \begin{bmatrix}
			\bm{I} & -\bm{C}\\
			-\bm{P} & \bm{I} \\
		\end{bmatrix}
	\end{align*}
	is causally invertible on $\mathcal{L}_{2e}^{n}\times \mathcal{L}_{2e}^n $.
\end{definition}
\begin{definition}[Closed-loop Stability]\label{def:closed_stable}
	A well-posed closed-loop system $\nlcls$ in Fig.~\ref{figloop} is said to be (finite-gain) stable if $\bm{F}^{-1}_{\bm{P},\bm{C}}$ is finite-gain stable.
\end{definition}

The following definition is tailored from \cite{angeli2006CCW}. 
\begin{definition}[Counterclockwise Dynamics]\label{def:CCW}
		A system $\bmP=u\mapsto y\in\mathcal{N}^{\mathcal{C}}_n$ is said to have counterclockwise (CCW) (input-output) dynamics if for any $T\geq 0~\text{and}~u\in\mathcal{L}_{2e}^n$, it holds
		$$\int_{0}^T \langle u(t),\dot{y}(t)\rangle dt\geq 0. $$
		It is said to have strictly CCW dynamics if for any $T\geq 0~\text{and}~u\in\mathcal{L}_{2e}^n$, it holds
		$$\int_{0}^T \left(\langle u(t),\dot{y}(t)\rangle - \frac{\rho(|\dot{y}(t)|)}{1+\gamma_\infty(|x(t)|)} \right)dt \geq 0,$$
		for some positive definite function $\rho$ and some strictly increasing function $\gamma_\infty$ with $\gamma_\infty(0)=0$ and $\lim_{a\to\infty}\gamma_\infty(a)=\infty$, where $x(t)$ is the state of the system.
\end{definition}

\section{Nonlinear Negative Imaginariness}\label{sec:NI}
In what follows, we present our definition for negative imaginary systems, and compare it with the well-received CCW dynamics introduced in the last subsection. 
\begin{definition}[Negative Imaginariness]\label{def:NI}
	A system $\bmP=u\mapsto y\in\mathcal{N}_n$ is said to be negative imaginary (NI)
	if \black{it is stable} and there exist $\bar{\Omega}^*\geq \underline{\Omega}^*>0$ such that for all $\bar{\Omega}\in[\bar{\Omega}^*,\infty)$ and $\underline{\Omega}\in(0,\underline{\Omega}^*]$, $$\Re\left(\int_{\underline{\Omega}}^{\bar{\Omega}}\langle \hat{u}(j\omega),j\omega \hat{y}(j\omega) \rangle d\omega\right) \geq 0,~\forall~u\in\mathcal{L}_2.$$
	It is said to be strictly negative imaginary (SNI) if \black{it is stable} and there exist $\bar{\Omega}^*\geq \underline{\Omega}^*>0$ such that for all $\bar{\Omega}\in[\bar{\Omega}^*,\infty)$ and $\underline{\Omega}\in(0,\underline{\Omega}^*]$, there exists $\epsilon=\epsilon(\bar{\Omega},\underline{\Omega})>0$ such that
	\black{\begin{multline*}
	\Re\left(\int_{\underline{\Omega}}^{\bar{\Omega}}\langle \hat{u}(j\omega),j\omega \hat{y}(j\omega) \rangle d\omega\right) \geq \\
	 \epsilon\int_{\underline{\Omega}}^{\bar{\Omega}}|\hat{u}(j\omega)|^2d\omega,~\forall~u\in\mathcal{L}_2.\end{multline*}}
\end{definition}
It is noteworthy that the integrals are taken only over finite frequency ranges that exclude $0$ and $\infty$, similarly to linear NI systems. \black{For notational simplicity, we say a statement holds for sufficiently large $\bar{\Omega}>0$ if there exists $\bar{\Omega}^*>0$ such that the statement holds for all $\bar{\Omega}\geq\bar{\Omega}^*$; similar convention applies to {sufficiently small} $\underline{\Omega}>0$ as well. }

\black{Negative imaginary systems form a convex cone, as is detailed in the following proposition. 
\begin{proposition}\label{prop:convex_NI}
	For any $\bm{G},\bm{H},\bm{K}\in\mathcal{N}_n$ with $\bm{G},\bm{H}$ being NI and $\bm{K}$ being SNI, it holds the following statements. \\
(a)~For $\alpha,\beta\geq 0$, $\alpha\bm{G}+\beta\bm{H}$ is NI. \\
(b)~For $\alpha\geq 0$ and $\beta>0$, 
		$\alpha\bm{G}+\beta\bm{K}$ is SNI. 
\end{proposition}
\begin{pf}
The two statements can be shown using similar arguments and  we only prove statement~(b) in what follows. 

Since $\bm{G}=u\mapsto y_1$ is NI, there exist $\bar{\Omega}_1^*\geq \underline{\Omega}_1^*>0$ such that for all $\bar{\Omega}\in[\bar{\Omega}_1^*,\infty)$ and $\underline{\Omega}\in(0,\underline{\Omega}_1^*]$, \begin{align}\label{eq:pf_prop_cone1}\Re\left(\int_{\underline{\Omega}}^{\bar{\Omega}}\langle \hat{u}(j\omega),j\omega \hat{y}_1(j\omega) \rangle d\omega\right) \geq 0,~\forall~u\in\mathcal{L}_2.\end{align}
Since $\bm{K}=u\mapsto y_2$ is SNI, there exist $\bar{\Omega}_2^*\geq \underline{\Omega}_2^*>0$ such that for all $\bar{\Omega}\in[\bar{\Omega}_2^*,\infty)$ and $\underline{\Omega}\in(0,\underline{\Omega}_2^*]$, there exists $\epsilon_2=\epsilon_2(\bar{\Omega},\underline{\Omega})>0$ such that
\begin{multline}\label{eq:pf_prop_cone2}
	\Re\left(\int_{\underline{\Omega}}^{\bar{\Omega}}\langle \hat{u}(j\omega),j\omega \hat{y}_2(j\omega) \rangle d\omega\right) \geq \\
	\epsilon_2\int_{\underline{\Omega}}^{\bar{\Omega}}|\hat{u}(j\omega)|^2d\omega,~\forall~u\in\mathcal{L}_2.\end{multline}
Using $\alpha\times \eqref{eq:pf_prop_cone1}+\beta\times\eqref{eq:pf_prop_cone2}$, we obtain that $\alpha\bm{G}+\beta\bm{K}=u\mapsto\alpha y_1+\beta y_2$ is SNI with $\bar{\Omega}^*=\max\{\bar{\Omega}^*_1,\bar{\Omega}^*_2\}$, $\underline{\Omega}^*=\min\{\underline{\Omega}^*_1,\underline{\Omega}^*_2\}$, and $\epsilon=\beta\epsilon_2(\bar{\Omega},\underline{\Omega})>0$.\hspace*{\fill}\qed 
\end{pf}}
\subsection{Relations to Systems with CCW Dynamics}
The proposed definition on nonlinear NI systems has a close relation to the well-studied notion of CCW dynamics, as revealed in the following lemmas. 

\begin{lemma}\label{lem:NI_CCW}
	Let $\bmP\in\mathcal{N}^{\mathcal{C}}_n$ be NI. Then $\bmP$ has CCW input-output dynamics. 
\end{lemma}
\begin{pf}
	By letting $\underline{\Omega}\to 0_+$ and $\bar{\Omega}\to\infty$, we obtain from Definition~\ref{def:NI} that for $u\in\mathcal{L}_2^n$
	\begin{multline}\label{eq:pf_lem1}
	0\leq \Re\left(\int_{\underline{\Omega}}^{\bar{\Omega}}\langle \hat{u}(j\omega),j\omega \hat{y}(j\omega) \rangle d\omega\right) \\
	\to \Re\left(\int_{0}^{\infty}\langle \hat{u}(j\omega),j\omega \hat{y}(j\omega) \rangle d\omega\right)\\
	=\dfrac{1}{2}\int_0^\infty \langle u(t),\dot{y}(t)\rangle dt\\
	 = \dfrac{1}{2} \int_0^\infty \left\langle u(t),\frac{d}{dt}(\bmP u)(t)\right\rangle dt. \end{multline}
	\black{Since $\bmP$ is causal, substituting $u$ with $\bmG_T \tilde{u}$ for all $\tilde{u}\in\mathcal{L}_{2e}^n$ and $T>0$ in the above inequality yields that
	\begin{multline*}\int_0^T \left\langle \tilde{u}(t),\frac{d}{dt}(\bmP \tilde{u})(t)\right\rangle dt \\
		= \int_0^\infty \left\langle (\bmG_T \tilde{u})(t),\frac{d}{dt}(\bmP \bmG_T \tilde{u})(t)\right\rangle dt\geq 0,\end{multline*}
	which completes the proof.} \hspace*{\fill}\qed
\end{pf}

{\begin{assumption}\label{ass:CCW}
	For all $u\in\mathcal{L}^n_2$ with a support of nonzero measure, it holds that $\dot{y}\in\mathcal{L}^n_2$ and
	\begin{align*}
		\int_0^\infty \langle u(t),\dot{y}(t)\rangle dt>0,
	\end{align*} 
where $y=\bmP u$.
\end{assumption}
\begin{lemma}\label{lem:CCW_NI}
	Let $\bmP\in\mathcal{N}^{\mathcal{C}}_n$ be stable, have CCW dynamics and satisfy Assumption~\ref{ass:CCW}.
	Then $\bmP$ is NI. 
\end{lemma}}
\begin{pf}
	Let $u\in\mathcal{L}^n_2$ with a support of nonzero measure. By Definition~\ref{def:CCW} and Plancherel's theorem, we have
	$$2\Re \left(\int_{0}^{\infty}\langle \hat{u}(j\omega),j\omega \hat{y}(j\omega) \rangle d\omega\right) =  \int_0^\infty \langle u(t),\dot{y}(t)\rangle dt > 0. $$
	Then for sufficiently large $\bar{\Omega}>0$ and sufficiently small $\underline{\Omega}>0$, it holds that for every $u\in\mathcal{L}^n_2$ with a support of nonzero measure, 
	$$\Re\left(\int_{\underline{\Omega}}^{\bar{\Omega}}\langle \hat{u}(j\omega),j\omega \hat{y}(j\omega) \rangle d\omega\right) \geq 0,$$
	which also holds for $u\in\mathcal{L}^n_2$ that is zero almost everywhere. \hspace*{\fill}\qed
\end{pf}

{Assumption~\ref{ass:CCW} can be weakened by strengthening $\bmP$ to have strictly CCW dynamics, yielding the following result.}
\begin{assumption}\label{ass:strictCCW}
	For all $u\in\mathcal{L}^n_2$ with a support of nonzero measure, it holds that $\dot{y}\in\mathcal{L}^n_2$ and has a support of nonzero measure, where $y=\bmP u$.
\end{assumption}
\begin{lemma}\label{lem:strictCCW_NI}
	Let $\bmP\in\mathcal{N}^{\mathcal{C}}_n$ be stable, have strictly CCW dynamics and satisfy Assumption~\ref{ass:strictCCW}. Then $\bmP$ is NI. 
\end{lemma}
\begin{pf}
	\black{For any $u\in\mathcal{L}^n_2$ with a support of nonzero measure and $y=\bmP u$, we have $\dot{y}\in\mathcal{L}^n_2$ and is nonzero for some time interval. Since $\bmP$ has strictly CCW input-output dynamics, we obtain from Definition~\ref{def:CCW} that
	$$\int_{0}^\infty \langle u(t),\dot{y}(t)\rangle dt \geq \int_{0}^\infty\frac{\rho(|\dot{y}(t)|)}{1+\gamma_\infty(|x(t)|)}>0,$$
	where the last inequality follows from that $\rho(\cdot)$ is positive definite and $\gamma_\infty(\cdot)$ is nonnegative.
	An application of Lemma~\ref{lem:CCW_NI} shows that $\bmP$ is NI.} \hspace*{\fill}\qed
\end{pf}

The above lemmas establish the relations {between} NI systems and those with CCW dynamics. In particular, Lemma~\ref{lem:NI_CCW} shows that NI systems (with differentiable outputs) necessarily have CCW dynamics, while Lemma~\ref{lem:CCW_NI} shows that under an additional positivity condition (in Assumption~\ref{ass:CCW}), a stable system with CCW dynamics is NI. Although we have seen that the proposed definition of nonlinear NI systems has a close relation to the systems with CCW dynamics, they are still essentially two different concepts. One can easily verify that a system with non-differentiable outputs cannot have CCW dynamics but it can be NI. 
Based on Lemma~\ref{lem:CCW_NI}, a system (satisfying Assumption~\ref{ass:strictCCW}) with strictly CCW is necessarily NI as shown in Lemma~\ref{lem:strictCCW_NI}, but not vice versa as will be demonstrated in Section~\ref{sec:sim}.

\subsection{Reduction to The LTI Case}
The following proposition shows that for LTI systems, Definition~\ref{def:NI} is equivalent to its original version in Definition~\ref{def:LTI_NI}.
\begin{proposition}\label{prop:NI_LTI}
	For an LTI $\bm{G}\in\mathcal{N}_n$ with transfer function $\hat{G}\in\mathcal{RH}_\infty^{n\times n}$, it holds the following relations. 
	
	(a) ~$\bm{G}$ is NI per Definition~\ref{def:NI} if and only if $$j(\hat{G}(j\omega)-\hat{G}(j\omega)^*)\geq 0,~\forall \omega\in(0,\infty).$$
	(b) ~$\bm{G}$ is SNI per Definition~\ref{def:NI} if and only if $$j(\hat{G}(j\omega)-\hat{G}(j\omega)^*)>0,~\forall \omega\in(0,\infty).$$	
\end{proposition}
\begin{pf}
	We start with statement (b). First we show sufficiency. By the conjugate symmetric property over imaginary axis of real-rational transfer matrices, {i.e., $\hat{G}(j\omega) = \overline{\hat{G}(-j\omega)}$}, we have for all $\bar{\Omega}>\underline{\Omega}>0$, it holds for $u\in\mathcal{L}_2^n$ that
	\begin{align}\label{eq:pf_prop1}
			&2\Re\left(\int_{\underline{\Omega}}^{\bar{\Omega}} \langle \hat{u}(j\omega),j\omega \hat{y}(j\omega) \rangle d\omega\right)\nonumber\\
			&= \int_{\underline{\Omega}}^{\bar{\Omega}} \langle \hat{u}(j\omega),j\omega (\hat{G}(j\omega)-\hat{G}(j\omega)^*)\hat{u}(j\omega) \rangle d\omega\\
			&\geq \min_{\omega\in[\underline{\Omega},\bar{\Omega}]}\underline{\sigma}(j\omega(\hat{G}(j\omega)-\hat{G}(j\omega)^*))\int_{\underline{\Omega}}^{\bar{\Omega}}|\hat{u}(j\omega)|^2d\omega.\nonumber
	\end{align}
	Sufficiency is thus shown by the last inequality with $$\epsilon:=\dfrac{1}{2}\min_{\omega\in[\underline{\Omega},\bar{\Omega}]}\underline{\sigma}(j\omega(\hat{G}(j\omega)-\hat{G}(j\omega)^*))>0.$$
	
	\black{Next we show necessity. For each $\hat{\omega}>0$, let $0<\Omega_a\leq \hat{\omega}\leq\Omega_b$ with $\Omega_b-\Omega_a=\eta>0$ being sufficiently small.  
	Construct that 
	\begin{align}\label{eq:pf_lem1_uhat}
		\hat{u}(j\omega)=\left\{\begin{matrix}\eta^{-1/2}\bar{v},&~~~~~\omega\in[-\Omega_b,-\Omega_a]\\ \eta^{-1/2}{v},&\omega\in[\Omega_a,\Omega_b]\\0,&\text{otherwise~~~}\end{matrix}\right.\end{align}
	where $v\in\mathbb{C}^n$ is an eigenvector of unit length corresponding to the least eigenvalue (also the least singular value) of $j\hat{\omega}(\hat{G}(j\hat{\omega})-\hat{G}(j\hat{\omega})^*)$. 
	It then follows from the definition of SNI that there exist $(\underline{\Omega},\bar{\Omega})\supset(\Omega_a,\Omega_b)$ and $\epsilon>0$ such that
	\begin{equation}\label{eq:pf_prop1_nec}
	\begin{aligned}
		&\int_{\Omega_a}^{\Omega_b} \langle \hat{u}(j\omega),j\omega (\hat{G}(j\omega)-\hat{G}(j\omega)^*)\hat{u}(j\omega) \rangle d\omega\\
		&=2\Re\int_{\underline{\Omega}}^{\bar{\Omega}}  \langle \hat{u}(j\omega),j\omega \hat{G}(j\omega)\hat{u}(j\omega) \rangle d\omega\\
		&\geq 2\epsilon\int_{\underline{\Omega}}^{\bar{\Omega}} |\hat{u}(j\omega)|^2d\omega =2\epsilon>0.
	\end{aligned}
\end{equation}
	Since $\hat{G}\in\mathcal{RH}_\infty$, $\hat{G}(j\omega)$ is continuous in $\omega\in\mathbb{R}$,
	whereby 
	$$0<2\epsilon\leq \text{LHS of \eqref{eq:pf_prop1_nec}} \to \sigma_n(j\hat{\omega}(\hat{G}(j\hat{\omega})-\hat{G}(j\hat{\omega})^*))$$ as $\eta\to 0_+$. Then by arbitrariness of $\hat{\omega}>0$, we conclude that
	$\sigma_n(j{\omega}(\hat{G}(j{\omega})-\hat{G}(j{\omega})^*))>0,~\forall \omega>0$.}
	
	\black{For statement (a), its sufficiency can be shown by using \eqref{eq:pf_prop1}. For the necessity part, follow the similar notation, and we complete the proof by noting that for $\hat{\omega}\in[\Omega_a,\Omega_b]$ and $(\underline{\Omega},\bar{\Omega})\supset(\Omega_a,\Omega_b)$, it holds
	\begin{align*}
		0&\leq 2\Re\int_{\underline{\Omega}}^{\bar{\Omega}}  \langle \hat{u}(j\omega),j\omega \hat{G}(j\omega)\hat{u}(j\omega) \rangle d\omega\\
		&=\int_{\Omega_a}^{\Omega_b} \langle \hat{u}(j\omega),j\omega (\hat{G}(j\omega)-\hat{G}(j\omega)^*)\hat{u}(j\omega) \rangle d\omega\\
		&\to\sigma_n(j\hat{\omega}(\hat{G}(j\hat{\omega})-\hat{G}(j\hat{\omega})^*))
	\end{align*}
	as $\eta=\Omega_b-\Omega_a \to 0_+$.}  \hspace*{\fill}\qed
\end{pf}

\begin{remark}
{As revealed by Lemma~\ref{lem:CCW_NI}, a nonlinear NI system can be identified by verifying if it possesses CCW dynamics. In addition, by Propositions~\ref{prop:convex_NI} and \ref{prop:NI_LTI}, a special class of nonlinear SNI systems can be identified, based on a given nonlinear NI system $\bm{G}$, as
\begin{multline*}
	\big\{\alpha\bm{G}+\beta\bm{H}~|~\alpha\geq0,~\beta>0,\\
	j(\hat{H}(j\omega)-\hat{H}(j\omega)^*)>0,~\forall~\omega\in(0,\infty)\big\},
\end{multline*}
where $\hat{H}$ is the transfer matrix characterizing an LTI system $\bm{H}$. }
\end{remark}

\section{Main Result}\label{sec:main}
In this section, we introduce our main results for the feedback stability of (nonlinear) NI systems. 

We start with the following proposition for feedback stability presented in terms of finite-frequency IQCs, which are adopted to handle the singularity frequencies around zero and infinity. 
\begin{proposition}\label{prop:main1}
	Let $\bmP,\bmC\in\mathcal{N}_n$, $\bmP$ be SNI, $\bmC$ be NI, and $\bmP\,\#\,(\tau\bmC)$ be well-posed for $\tau\in[0,1]$. Suppose that there exist $\bar{\Omega}^*>\underline{\Omega}^*>0$, $\epsilon_0,\epsilon_\infty>0$, and $\Pi_0, \Pi_\infty\in\mathcal{L}^{2n\times 2n}_\infty$ with $\Pi_0(j\omega)^*=\Pi_0(j\omega)$ and $\Pi_\infty(j\omega)^*=\Pi_\infty(j\omega)$, $\omega\in\mathbb{R}$, such that for all $\underline{\Omega}\in(0,\underline{\Omega}^*]$, $\bar{\Omega}\in[\bar{\Omega}^*,\infty)$, $\tau\in[0,1]$, $u,y\in\mathcal{L}^n_2$, it holds
	\begin{equation}\label{eq:prop_IQC}
	\begin{aligned}
	&\Re\left(\int_0^{\underline{\Omega}} \begin{bmatrix}(\widehat{\bmP u}) \\ \hat{u}\end{bmatrix}^*\Pi_0 \begin{bmatrix}(\widehat{\bmP u}) \\ \hat{u}\end{bmatrix}d\omega \right)\leq -\epsilon_0 \int_0^{\underline{\Omega}}|\hat{u}|^2d\omega,\\
	&\Re\left(\int_0^{\underline{\Omega}} \begin{bmatrix} \hat{y} \\ \widehat{(\tau\bmC y)}\end{bmatrix}^*\Pi_0 \begin{bmatrix} \hat{y} \\ \widehat{(\tau\bmC y)}\end{bmatrix}d\omega\right) \geq 0,\\
	&\Re\left(\int_{\Bar{\Omega}}^\infty \begin{bmatrix}(\widehat{\bmP u}) \\ \hat{u}\end{bmatrix}^*\Pi_\infty \begin{bmatrix}(\widehat{\bmP u}) \\ \hat{u}\end{bmatrix}d\omega\right) \leq -\epsilon_\infty \int_{\bar{\Omega}}^\infty|\hat{u}|^2d\omega,\\
	&\Re\left(\int_{\Bar{\Omega}}^\infty \begin{bmatrix} \hat{y} \\ \widehat{(\tau\bmC y)}\end{bmatrix}^*\Pi_\infty \begin{bmatrix} \hat{y} \\ \widehat{(\tau\bmC y)}\end{bmatrix}d\omega\right) \geq 0.
	\end{aligned}
	\end{equation}
	Then $\nlcls$ is stable.
\end{proposition}
\begin{pf}
	Denote by
	$$\Pi_m(\omega):=\begin{bmatrix}0 & j\omega I_n\\ -j\omega I_n & 0\end{bmatrix}.$$
	By Definition~\ref{def:NI}, $\tau\bmC$ is NI for all $\tau\in[0,1]$. Together with that $\bmP$ is SNI, we obtain that there exist $\underline{\Omega}\in(0,\underline{\Omega}^*]$, $\bar{\Omega}\in[\bar{\Omega}^*,\infty)$  and $\epsilon_m>0$ such that for all $u,y\in\mathcal{L}_2^n$,
	\begin{equation}\label{eq:pf2}
	\begin{aligned}
	&\Re\left(\int_{\underline{\Omega}}^{\bar{\Omega}} \begin{bmatrix}(\widehat{\bmP u}) \\ \hat{u}\end{bmatrix}^*\Pi_m \begin{bmatrix}(\widehat{\bmP u}) \\ \hat{u}\end{bmatrix}d\omega\right) \leq -\epsilon_m \int_{\underline{\Omega}}^{\bar{\Omega}}|\hat{u}|^2d\omega,\\
	&\Re\left(\int_{\underline{\Omega}}^{\bar{\Omega}} \begin{bmatrix} \hat{y} \\ \widehat{(\tau\bmC y)}\end{bmatrix}^*\Pi_m \begin{bmatrix} \hat{y} \\ \widehat{(\tau\bmC y)}\end{bmatrix}d\omega\right) \geq 0.
	\end{aligned}
	\end{equation}
	Define that $\Pi(\omega):=\gamma_m(\omega)\Pi_m+\gamma_0(\omega)\Pi_0+\gamma_\infty(\omega)\Pi_\infty$, where
	\begin{multline*}\gamma_0(\omega)=\left\{\begin{matrix}1,&|\omega|\leq\underline{\Omega}\\  0,&\text{otherwise}\end{matrix}\right.,~\gamma_\infty(\omega)=\left\{\begin{matrix}1,&|\omega|\geq\bar{\Omega}\\
	0,&\text{otherwise}\end{matrix}\right.,\\
	~\text{and}~\gamma_m(\omega)=1-\gamma_0(\omega)-\gamma_\infty(\omega).\end{multline*}
	It can be easily verified that $\Pi\in\mathcal{L}_\infty$. 
	Let $\epsilon:=\min\{\epsilon_0,\epsilon_\infty,\epsilon_m\}>0$. By combining \eqref{eq:prop_IQC} and \eqref{eq:pf2} properly,	
	we obtain that
	\begin{equation*}
	\begin{aligned}
	&\Re\left(\int_{0}^{\infty} \begin{bmatrix}(\widehat{\bmP u}) \\ \hat{u}\end{bmatrix}^*\Pi \begin{bmatrix}(\widehat{\bmP u}) \\ \hat{u}\end{bmatrix}d\omega\right) \leq -\epsilon\int_{0}^{\infty}|\hat{u}|^2d\omega\\
	&\hspace{60pt}\leq -\frac{\epsilon}{1+\|\bmP\|^2}\int_{0}^{\infty}\left|\begin{bmatrix}(\widehat{\bmP u}) \\ \hat{u}\end{bmatrix}\right|^2d\omega,\\
	&\Re\left(\int_{0}^\infty \begin{bmatrix} \hat{y} \\ \widehat{(\tau\bmC y)}\end{bmatrix}^*\Pi \begin{bmatrix} \hat{y} \\ \widehat{(\tau\bmC y)}\end{bmatrix}d\omega\right) \geq 0,
	\end{aligned}
	\end{equation*}
	where the second inequality follows by Lemma~\ref{lem:finite_gain_stability_def}, namely, for $u\in\mathcal{L}_2^n$, \begin{multline*} 
		2\int_{0}^{\infty}|\widehat{(\bmP u)}|^2d\omega =\|\bmP u\|_2^2\leq \|\bmP\|^{2}\|u\|^2_2\\ =  2\|\bmP\|^{2}\int_{0}^{\infty}|\hat{u}|^2d\omega.
	\end{multline*}%
	Note that $\bmP\,\#\,(\tau\bmC)$ is well-posed for $\tau\in[0,1]$. The stability of $\nlcls$ is then established in light of \cite[Corollary~IV.3]{khong2021IQC}, \cite[Theorem~2]{rantzer1997integral}.\hspace*{\fill}\qed
\end{pf}

It is noteworthy that Proposition~\ref{prop:main1} is an extension of \cite[Theorem~4]{KhongPR17} and is the first stability result involving nonlinear NI systems defined in Definition~\ref{def:NI}. It generalizes some versions of the previously studied feedback stability results on linear NI systems in \cite{Lanzon2008TAC,KhongPR17}. However, verifying the inequalities in \eqref{eq:prop_IQC} for nonlinear systems is much more difficult than their counterparts in the LTI setting. \black{As a result, the inequalities will be further interpreted in what follows so as to obtain more concise and verifiable feedback stability results for nonlinear NI systems. }

A class of nonlinear systems, which satisfies a certain IQC on the time-averaged input-output signals, is defined as
\begin{align*}
&\mathcal{B}(\Xi,\epsilon):=\left\{\bmP\in\mathcal{N}_n~\Bigg|~\text{for all}~u\in\mathcal{L}_1\cap\mathcal{L}_2~\text{such that}\right.\\
&~y=\bmP u\in\mathcal{L}_1\cap\mathcal{L}_2,
\begin{bmatrix}\bar{y}\\\bar{u}\end{bmatrix}^T\Xi\begin{bmatrix}\bar{y}\\\bar{u}\end{bmatrix}\leq \black{-\epsilon|\bar{u}|^2},~\text{where}\\
&\left.\bar{u}:=\int_0^\infty u(t)dt,~\bar{y}:=\int_0^\infty y(t)dt, 
~\right\},
\end{align*}
for $\Xi\in\mathbb{C}^{2n\times 2n}$ with $\Xi^*=\Xi$ and $\epsilon\geq 0$. In addition, we define a complementary set via that $\bmP\in\mathcal{B}_C(\Xi,\epsilon)$ if $\bmP\in\mathcal{B}(\tilde{\Xi},\epsilon)$, 
where $$\tilde{\Xi}=-\begin{bmatrix}0 & I_n \\ I_n& 0 \end{bmatrix}\Xi\begin{bmatrix}0 & I_n \\ I_n& 0 \end{bmatrix}.$$
\black{In the above sets, we restrict the signals to be in $\mathcal{L}_1$ so as to ensure that $\bar{u}$ and $\bar{y}$ are well defined.}

{A special class of systems that belong to the above sets is given as follows. 
\begin{lemma}\label{lem:setofB}
An LTI system $\bm{G}\in\mathcal{N}_n$ with transfer matrix $\hat{G}\in\mathcal{RH}_\infty^{n\times n}$ belongs to $\mathcal{B}(\Xi,\epsilon)$ if 
\begin{align}\label{eq:lemB}
	\begin{bmatrix}\hat{G}(j0)\\I_n\end{bmatrix}^T\Xi\begin{bmatrix}\hat{G}(j0) \\ I_n\end{bmatrix}\leq -\epsilon I_n.
\end{align}
\end{lemma}
\begin{pf}
	Let $u\in\mathcal{L}_1\cap\mathcal{L}_2$ such that $y=\bm{G}u\in\mathcal{L}_1\cap\mathcal{L}_2$. It then follows that $\hat{y}=\hat{G}\hat{u}$, whereby $\hat{y}(j0) = \hat{G}(j0)\hat{u}(j0)$. By the definition of Fourier transform, we obtain $\bar{y} = \hat{G}(j0) \bar{u}$, where
	$$\bar{u}=\int_0^\infty u(t)dt,~\bar{y}=\int_0^\infty y(t)dt.$$
	Multiplying $\bar{u}^T$ and $\bar{u}$ at both sides of \eqref{eq:lemB} yields that
	\begin{align*}
	-\epsilon |\bar{u}|^2 \geq \begin{bmatrix}\hat{G}(j0)\bar{u}\\\bar{u}\end{bmatrix}^T\Xi\begin{bmatrix}\hat{G}(j0)\bar{u} \\ \bar{u}\end{bmatrix}=\begin{bmatrix}\bar{y}\\ \bar{u}\end{bmatrix}^T\Xi\begin{bmatrix}\bar{y}\\ \bar{u}\end{bmatrix},
	\end{align*}
which completes the proof.\hspace*{\fill}\qed
\end{pf}
}
The uniform instantaneous gain of $\bmP\in\mathcal{N}_n$ that is locally Lipschitz continuous is defined as \cite[Chapter~4]{Willems1971nonlinear} 
\begin{multline*}\gamma(\bmP):=\sup_{T\geq 0}\lim_{\Delta_T\to 0_+}\sup_{x,y}\left\{\frac{\|\bmG_{T+\Delta_T}(\bmP y-\bmP x)\|_2}{\|\bmG_{T+\Delta_T}(y-x)\|_2}\bigg|\right.\\
\left. x,y\in\mathcal{L}^n_2,~\bmG_{T}( y- x)=0,~\bmG_{T+\Delta_T}( y- x)\neq 0\right\}.\end{multline*}

We are now ready to present our main stability result in the following theorem. Its proof is deferred to Section~\ref{subsec:proof}.
\begin{theorem}\label{thm:SNI2}
	Let $\bmP,\bmC\in\mathcal{N}_n$ be both locally Lipschitz continuous with $\gamma(\bmP)<\alpha$ and $\gamma(\bmC)<\alpha^{-1}$ for some $\alpha>0$. Then $\bmP\,\#\,\bmC$ is well-posed and stable, if there exist $\Xi=\Xi^*\in\mathbb{C}^{2n\times 2n}$ and $\epsilon>0$ such that $\bmP$ is SNI with $\bmP\in\mathcal{B}(\Xi,\epsilon)$ and $\bmC$ is NI with $\tau\bmC\in\mathcal{B}_C(\Xi,0)$, $\forall~\tau\in[0,1]$.  
\end{theorem}
\begin{remark}
	It is noteworthy that if either of $\gamma(\bmP)$ and $\gamma(\bmC)$ is zero, the other can take an arbitrary value in the above theorem. Moreover, in combination with Proposition~\ref{prop:NI_LTI}, we have that both Proposition~\ref{prop:main1} and Theorem~\ref{thm:SNI2} reduce to \cite[Theorem~4]{KhongPR17} when $\bmP$ and $\bmC$ are taken to be LTI. 
\end{remark}
The following is a typical class of nonlinear systems that are NI and satisfy the IQC constraints in Theorem~\ref{thm:SNI2}.
\begin{example}\label{ex:NI}
	Consider the following input-to-state stable (ISS) \cite[Chapter~5]{khalil1996nonlinearbook} nonlinear system
	$$\bmP:~u\mapsto y\in\mathcal{N}_1~\text{with}~\left\{\begin{array}{l}\dot{x}_1=x_2-u\\
		\dot{x}_2=f(x,u)\\
		y=x_2\end{array}\right. ,$$
	where  $x(0)=0$ and $f:~\mathbb{R}^2\times\mathbb{R}\to\mathbb{R}$ is continuously differentiable with $f(0,0)=0$.
	Since $\bmP\in\mathcal{N}_1$ is ISS, for all $u\in\mathcal{L}_2$, it holds that $x\in\mathcal{L}_2$. Let $u,x\in\mathcal{L}_1\cap\mathcal{L}_2$, which is a dense subset in $\mathcal{L}_2$ \cite[Chapter~2]{Elliott2001Analysis}. It then holds that
	$$\int_0^\infty y(t)dt=x_1(\infty)-x_1(0)+\int_0^\infty u(t)dt=\int_0^\infty u(t)dt, $$
	yielding that $\bmP\in\mathcal{B}_C(\Xi,\epsilon)$ for all $\Xi=\Xi^*\in\mathbb{C}^{2\times 2}$ and $\epsilon \geq 0$ with the elements in $\Xi=\begin{bmatrix}
		\Xi_{11} & \Xi_{12} \\\Xi^*_{12} & \Xi_{22}
	\end{bmatrix}$ satisfying
	\begin{align}\label{eq:example}
	\black{\Xi_{11}+\Xi_{12}+\Xi^*_{12}+\Xi_{22}\geq \epsilon.}
	\end{align} 
	If we take $f(x,u)=-3x_1-x_2+u$, we can verify that $\bmP$ is LTI, ISS and SNI. Furthermore, if we take 
	$$f(x,u)=-3x_1-\frac{x_2}{1+x_2^2}+u$$ which is nonlinear, we can verify that $\bmP$ is ISS and for $\forall~u\in\mathcal{L}_2$ it holds
	$$\int_0^\infty u\dot{y}dt=\int_0^\infty\frac{(u-y)^2+u^2y^2}{1+y^2}dt \geq 0.$$
	{By Definition~\ref{def:CCW}, $\bmP$ has CCW dynamics. It then follows from Lemma~\ref{lem:CCW_NI} that $\bmP$ is NI.} 
\end{example}
\subsection{Proof of the Main Result}\label{subsec:proof}
\begin{lemma}\label{lem:high_freq}
	Let $\bmP\in\mathcal{N}_n$ be stable and locally Lipschitz continuous. It then holds that
	$$\sup_{u\in\mathcal{L}^n_2}\lim_{\omega\to\infty}\frac{|\widehat{(\bmP u)}(j\omega)|}{|\hat{u}(j\omega)|} \leq \gamma(\bmP).$$	
\end{lemma}
\begin{pf}
	By the definition of the uniform instantaneous gain, we obtain that
	\begin{align*}
		&\gamma(\bmP)\geq \lim_{\Delta_T\to 0_+}\sup_x\left\{\frac{\|\bmG_{\Delta_T} \bmP x\|_2}{\|\bmG_{\Delta_T}x\|_2}~\bigg|~x\in\mathcal{L}^n_2,~\bmG_{\Delta_T} x\neq 0\right\}\\
		&=\sup_x \lim_{\Delta_T\to 0_+}\hspace{-5pt}\left\{\frac{\sqrt{\int_0^{\Delta_T}| (\bmP x)(t)|^2dt}}{\sqrt{\int_0^{\Delta_T}| x(t)|^2dt}}\Bigg|x\in\mathcal{L}^n_2,\bmG_{\Delta_T} x\neq 0\right\}\\
		&=\sup_x \left\{\lim_{t\to 0_+}\frac{| (\bmP x)(t)|}{| x(t)|}~\bigg|~x\in\mathcal{L}^n_2, x(0)\neq 0\right\}\\
		&=\sup_x \left\{\lim_{\omega\to \infty}\frac{| j\omega\widehat{(\bmP x)}(j\omega)|}{|j\omega\hat{x}(j\omega)|}~\bigg|~x\in\mathcal{L}^n_2\right\}\\
		&= \sup_{x\in\mathcal{L}^n_2}\lim_{\omega\to\infty}\frac{|\widehat{(\bmP x)}(j\omega)|}{|\hat{x}(j\omega)|},
	\end{align*}
\black{where the second last equality follows by applying the initial value theorem \cite[Chapter~2]{fourier2017} on ${x}$ and ${\bmP x}$, respectively.} \hspace*{\fill}\qed
\end{pf}

Based on Proposition~\ref{prop:main1} and Lemma~\ref{lem:high_freq}, the proof of Theorem~\ref{thm:SNI2} is given below. 
\begin{pf}
	Since $\mathcal{L}_1^n\cap\mathcal{L}_2^n$ is dense in $\mathcal{L}_2^n$ \cite[Chapter~2]{Elliott2001Analysis}, and $\bmP$ and $\bmC$ are stable and locally Lipschitz continuous, it suffices to examine their input-output pairs in $\mathcal{L}_1^n\cap\mathcal{L}_2^n$ to show the feedback stability in what follows. 
	Note that for every $x\in\mathcal{L}_1^n\cap\mathcal{L}_2^n$, it follows from the definition of Fourier transform that
	$$\hat{x}(j0)={\frac{1}{\sqrt{2\pi}}}\int_0^\infty x(t)dt<\infty.$$
	Since $\bmP\in\mathcal{B}(\Xi,\epsilon)$, it holds for all $u\in\mathcal{L}_1^n\cap\mathcal{L}_2^n$ satisfying $\bmP u\in\mathcal{L}_1^n\cap\mathcal{L}_2^n$ that
	\begin{equation}\label{eq:pf_main}
	\begin{aligned}
	\lim_{\underline{\Omega}\to 0_+}& \frac{1}{\underline{\Omega}}\Re\int_{0}^{\underline{\Omega}}  \begin{bmatrix}(\widehat{\bmP u}) \\ \hat{u}\end{bmatrix}^*\Xi \begin{bmatrix}(\widehat{\bmP u}) \\ \hat{u}\end{bmatrix}d\omega\\
	&=\begin{bmatrix}(\widehat{\bmP u})(j0) \\ \hat{u}(j0)\end{bmatrix}^T\Xi \begin{bmatrix}(\widehat{\bmP u})(j0) \\ \hat{u}(j0)\end{bmatrix}\\
	&\leq -\epsilon|\hat{u}(j0)|^2=-\epsilon\lim_{\underline{\Omega}\to 0_+} \frac{1}{\underline{\Omega}}\Re\int_0^{\underline{\Omega}} |\hat{u}|^2d\omega.
	\end{aligned}
    \end{equation}
	Since $\tau\bmC\in\mathcal{B}_C(\Xi,0)$, $\tau\in[0,1]$, similarly to the above inequality we have for all $y\in\mathcal{L}_1^n\cap\mathcal{L}_2^n$ such that $\bmC y\in \mathcal{L}_1^n\cap\mathcal{L}_2^n$,
	\begin{multline*}
	\lim_{\underline{\Omega}\to 0_+} \frac{1}{\underline{\Omega}}\Re\int_{0}^{\underline{\Omega}} \begin{bmatrix} \hat{y} \\ \widehat{(\tau\bmC y)}\end{bmatrix}^*\Xi\begin{bmatrix} \hat{y} \\ \widehat{(\tau\bmC y)}\end{bmatrix}d\omega \\ =\begin{bmatrix} \hat{y}(j0) \\ \widehat{(\tau\bmC y)}(j0)\end{bmatrix}^T\Xi\begin{bmatrix} \hat{y}(j0) \\ \widehat{(\tau\bmC y)}(j0)\end{bmatrix}\geq 0.
	\end{multline*}
	Let $\epsilon_0=\epsilon/3>0$. \black{Using the fact that $\mathcal{L}_1^n\cap\mathcal{L}_2^n$ is dense in $\mathcal{L}_2^n$ and the premise that $\bmP$ and $\bmC$ are stable and locally Lipschitz continuous, we obtain that there exists $\underline{\Omega}^*>0$ such that for all $\underline{\Omega}\in(0,\underline{\Omega}^*]$ and $u,y\in\mathcal{L}_2^n$,}
	\begin{equation}\label{eq:pf_thm1_lowfreq}
	\begin{aligned}
	&\Re\int_0^{\underline{\Omega}} \begin{bmatrix}(\widehat{\bmP u}) \\ \hat{u}\end{bmatrix}^*\Pi_0 \begin{bmatrix}(\widehat{\bmP u}) \\ \hat{u}\end{bmatrix}d\omega\\
	&\hspace{100pt}\leq -\epsilon_0 \int_0^{\underline{\Omega}}|\hat{u}|^2d\omega,\\
	&\Re\int_0^{\underline{\Omega}} \begin{bmatrix} \hat{y} \\ \widehat{(\tau\bmC y)}\end{bmatrix}^*\Pi_0 \begin{bmatrix} \hat{y} \\ \widehat{(\tau\bmC y)}\end{bmatrix}d\omega \geq 0.
	\end{aligned}
	\end{equation}
	where \black{$$\Pi_0=\Xi+\epsilon_0\begin{bmatrix}0 & 0 \\ 0 & I_n\end{bmatrix}\in\mathbb{C}^{2n\times 2n}\subset\mathcal{L}_\infty.$$}
	
	By hypothesis, we have $\gamma(\bmP)<\alpha$ and $\gamma(\bmC)<\alpha^{-1}$. Then there exist $\epsilon_2,\epsilon_\infty>0$ such that
	$$\gamma^2(\bmC)<\alpha^{-2}-\epsilon_2~\text{and}~(\epsilon_\infty+\alpha^2)(\alpha^{-2}-\epsilon_2)<1.$$
	\black{Using Lemma~\ref{lem:high_freq}, we obtain that there exists $\bar{\Omega}^*>\underline{\Omega}^*$, such that for all $\bar{\Omega}\in[\bar{\Omega}^*,\infty)$, $\tau\in[0,1]$ and $u,y\in\mathcal{L}_2^n$,
	\begin{align*}
	&\int_{\bar{\Omega}}^\infty|\widehat{(\bmP u)}|^2d\omega \leq \alpha^2\int_{\bar{\Omega}}^\infty|\hat{u}|^2d\omega,\\
	&\int_{\bar{\Omega}}^\infty|\widehat{(\tau\bmC y)}|^2d\omega \leq (\alpha^{-2}-\epsilon_2)\int_{\bar{\Omega}}^\infty|\hat{y}|^2d\omega\\
	&\hspace{60pt}\leq(\epsilon_\infty+\alpha^2)^{-1}\int_{\bar{\Omega}}^\infty|\hat{y}|^2d\omega.
	\end{align*}
	Rearranging the above inequalities yields that
	\begin{equation}\label{eq:pf_thm1_high1}
	\begin{aligned}
	&\Re\left(\int_{\bar{\Omega}}^\infty \begin{bmatrix}(\widehat{\bmP u}) \\ \hat{u}\end{bmatrix}^*\Pi_\infty \begin{bmatrix}(\widehat{\bmP u}) \\ \hat{u}\end{bmatrix}d\omega \right)\\
	&\hspace{100pt}\leq -\epsilon_\infty \int_{\bar{\Omega}}^\infty|\hat{u}|^2d\omega,\\
	&\Re\left(\int_{\Bar{\Omega}}^\infty \begin{bmatrix} \hat{y} \\ \widehat{(\tau\bmC y)}\end{bmatrix}^*\Pi_\infty \begin{bmatrix} \hat{y} \\ \widehat{(\tau\bmC y)}\end{bmatrix}d\omega\right) \geq 0.
	\end{aligned}
	\end{equation}
	where \black{$$\Pi_\infty=\begin{bmatrix}I_n & 0 \\ 0 & -(\epsilon_\infty+\alpha^2)I_n\end{bmatrix}\in\mathbb{C}^{2n\times 2n}\subset\mathcal{L}_\infty.$$}
	
	By \cite[Chapter~4]{Willems1971nonlinear}, we obtain the well-posedness of $\bmP\,\#\,(\tau\bmC)$ from $\gamma(\bmP)\gamma(\bmC)<1$, $\tau\in[0,1]$. Therefore, using Proposition~\ref{prop:main1} with \eqref{eq:pf_thm1_lowfreq} and \eqref{eq:pf_thm1_high1}, we obtain the stability of $\bmP\,\#\,\bmC$.}\hspace*{\fill}\qed
\end{pf}

\subsection{Useful Corollaries}
The following corollary gives a sufficient condition for feedback stability between a linear SNI system and a nonlinear system with CCW dynamics. {An illustrative example for this corollary is provided in Section~\ref{subsec:eg_linear_nl}.}
{\begin{corollary}\label{cor:strictCCW_LTISNI}
		Let LTI system $\bmP$ with transfer matrix $\hat{P}\in\mathcal{RH}_\infty^{n\times n}$ satisfy that $j(\hat{P}(j\omega)-\hat{P}(j\omega)^*)>0,~\forall \omega>0$. Let $\bmC\in\mathcal{N}^{\mathcal{C}}_n$ be stable, satisfy Assumption~\ref{ass:CCW} and have CCW dynamics.
		Then $\bmP\;\#\;\bmC$ is well-posed and stable if there exists $\Xi=\Xi^*\in\mathbb{C}^{2n\times 2n}$ such that
		\begin{equation}\label{eq:cor1}\begin{aligned}
		\begin{bmatrix}\hat{P}(j0)\\I_n\end{bmatrix}^T\Xi\begin{bmatrix}\hat{P}(j0)\\I_n\end{bmatrix}<0,\end{aligned}\end{equation}
	and that $\bmC \in\mathcal{B}_C(\Xi,0)$ is locally Lipschitz continuous with $\gamma(\bmC)\bar{\sigma}(\hat{P}(j\infty))<1$. 
	\end{corollary}
	\begin{pf}
		By Proposition~\ref{prop:NI_LTI}, we know $\bmP$ is SNI. {Combining \eqref{eq:cor1} with Lemma~\ref{lem:setofB}}, we obtain that there exists $\epsilon>0$ such that $\bmP\in\mathcal{B}(\Xi,\epsilon)$. Since $\bmP$ is LTI, it is locally Lipschitz continuous with $\gamma(\bmP)=\bar{\sigma}(\hat{P}(j\infty))$. 
		On the other hand, we obtain from Lemma~\ref{lem:CCW_NI} that $\bmC$ is NI and so is $\tau\bmC$, $\tau\in[0,1]$. Consequently, it follows from Theorem~\ref{thm:SNI2} that $\bmP\;\#\;\bmC$ is well-posed and stable.\hspace*{\fill}\qed
\end{pf}
}

The following corollary gives a sufficient condition for feedback stability between linear SNI and NI systems, which is exactly \cite[Theorem~4]{KhongPR17}.
\begin{corollary}\label{eq:cor_linear}
Let LTI system $\bmP$ with transfer matrix $\hat{P}$ be SNI and LTI system $\bmC$ with transfer matrix $\hat{C}$ be NI. Suppose there exist $\Xi_0=\Xi_0^*\in\mathbb{C}^{2n\times 2n}$ and $\Xi_\infty=\Xi_\infty^*\in\mathbb{C}^{2n\times 2n}$ such that for all $\tau\in[0,1]$, 
\begin{equation}\label{eq:cor2}
\begin{aligned}
	\begin{bmatrix}\hat{P}(j0)\\I_n\end{bmatrix}^T\Xi_0\begin{bmatrix}\hat{P}(j0)\\I_n\end{bmatrix}<0,\\
	\begin{bmatrix}I_n \\ \tau\hat{C}(j0)\end{bmatrix}^T\Xi_0\begin{bmatrix}I_n \\ \tau\hat{C}(j0)\end{bmatrix}\geq 0,\\
	\begin{bmatrix}\hat{P}(j\infty)\\I_n\end{bmatrix}^T\Xi_\infty\begin{bmatrix}\hat{P}(j\infty)\\I_n\end{bmatrix}<0,\\
	\begin{bmatrix}I_n \\ \tau\hat{C}(j\infty)\end{bmatrix}^T\Xi_\infty\begin{bmatrix}I_n \\ \tau\hat{C}(j\infty)\end{bmatrix}\geq 0.
\end{aligned}
\end{equation}
\end{corollary}
\begin{pf}
	The corollary is shown using Proposition~\ref{prop:main1} with $\Pi_0=\Xi_0$ and $\Pi_\infty=\Xi_\infty$ and together with a similar argument to that in \eqref{eq:pf_main}.\hspace*{\fill}\qed
\end{pf}

\section{Simulation Results}\label{sec:sim}
In this section, we simulate the behaviours of feedback interconnections between (nonlinear) negative imaginary systems. 
Recall the nonlinear system $\bmP=u\mapsto y\in\mathcal{N}_1$ in Example~\ref{ex:NI} given by
\begin{align}\label{eq:P_Sim}
\left\{\begin{array}{l}\dot{x}_1=x_2-u\\
\dot{x}_2=-3x_1-\dfrac{x_2}{1+x_2^2}+u\\
y=x_2\end{array}\right.,
\end{align}
which is ISS and NI. 
It can be easily verified that the uniform instantaneous gain of $\bmP$ is zero, i.e. $\gamma(\bmP)=0$. Moreover, It follows from \eqref{eq:example} that $\tau \bmP\in\mathcal{B}_C(\Xi_i,0)$ for all $\tau\in[0,1]$ and $i=1,2$, where 
$$\Xi_1=\begin{bmatrix}0 & 1\\ 1& 0\end{bmatrix}~~~\text{and}~~~\Xi_2=\begin{bmatrix}1 & 0\\ 0& -1\end{bmatrix}.$$
\subsection{Feedback Connection of Nonlinear NI and Linear SNI Systems}\label{subsec:eg_linear_nl}
\begin{figure}[H]
	\centering
	\includegraphics[scale=0.4]{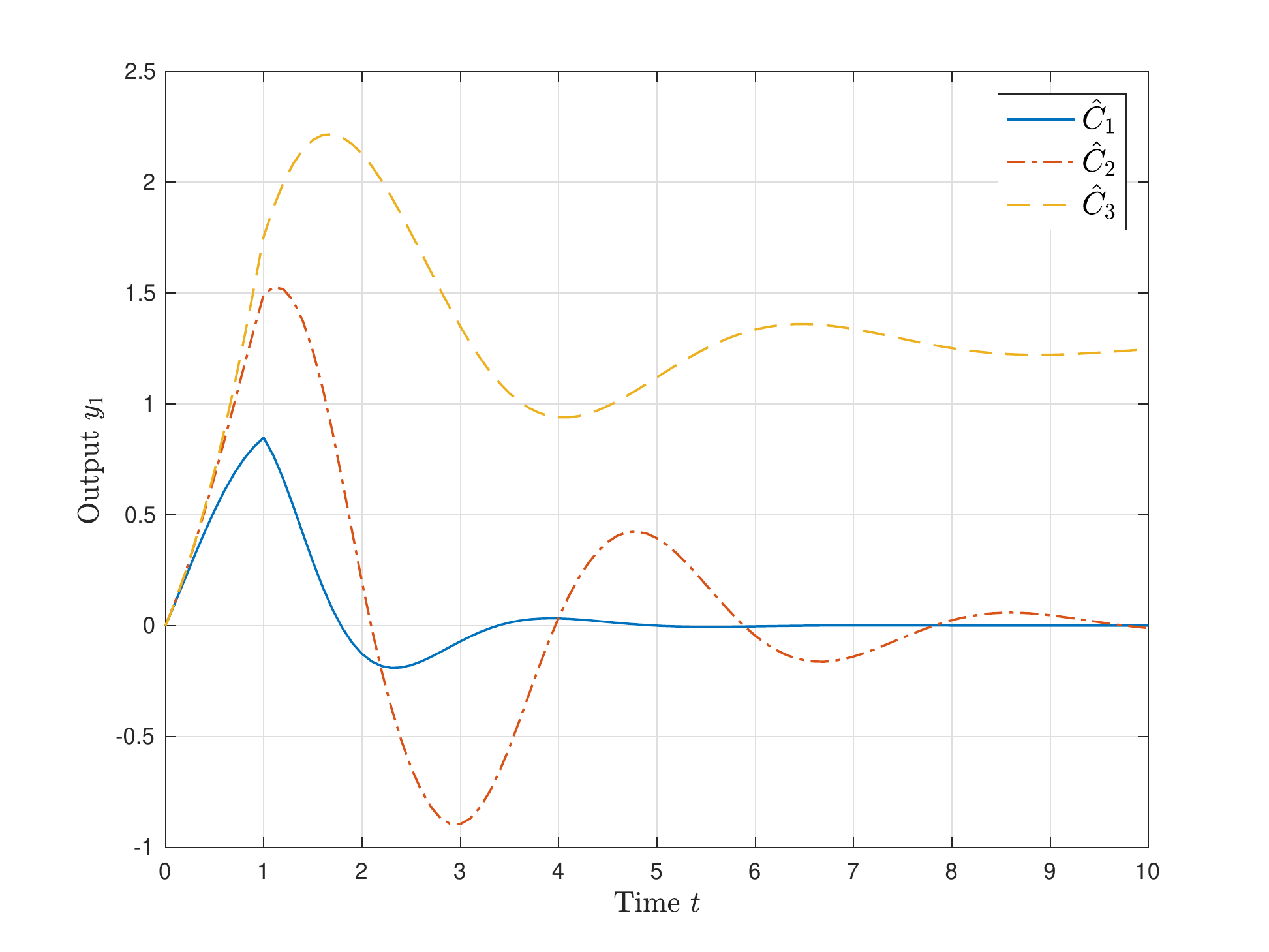}\\
	\caption{The output signals $y_1$ corresponding to different choices of $\bmC_i$, $i=1,2,3$, when $d_1$ is an impulse at time zero.}\label{figexampleNI}
\end{figure}
Given $\bmP$ in \eqref{eq:P_Sim}, consider the following candidates for $\bmC_i$ with $\hat{C}_i\in\mathcal{RH}_\infty$ that compose the feedback systems $\nlcls_i$, $i=1,2,3$:
$$\hat{C}_1=-\frac{s+2}{s+1},~\hat{C}_2=\frac{0.1}{s+1},~\text{and}~\hat{C}_3=\frac{1}{s+1}.$$
Clearly, all of them are linear SNI and {satisfy that 
$$\begin{bmatrix}\hat{C}_i(j0)\\I_n\end{bmatrix}^T\Xi_i\begin{bmatrix}\hat{C}_i(j0)\\I_n\end{bmatrix}<0,~i=1,2
$$
while
$$\begin{bmatrix}\hat{C}_3(j0)\\I_n\end{bmatrix}^T\Xi_i\begin{bmatrix}\hat{C}_3(j0)\\I_n\end{bmatrix}\nless 0,~i=1,2.$$}%
Actually, one can verify that there exists no $\Xi$ with $\Xi=\Xi^*$ and $\epsilon>0$ such that $\bmP\in\mathcal{B}_C(\Xi,0)$ and $\hat{C}_3\in\mathcal{B}(\Xi,\epsilon)$. The stability of $\nlcls_1$ and $\nlcls_2$ can be concluded from {Corollary~\ref{cor:strictCCW_LTISNI}} using $\Xi_1$ and $\Xi_2$, respectively. As shown in Fig.~\ref{figexampleNI}, $\nlcls_1$ and $\nlcls_2$ produce energy-bounded output signals given an impulse input signal, which is necessary for the feedback systems to be stable. On the other hand, note that $\nlcls_3$ does not satisfy the premises in Theorem~\ref{thm:SNI2}, and the simulation also reveals that such a system is unstable.

\subsection{Feedback Connection of Two Nonlinear Systems}\label{subsec:eg_nonlinear}
\begin{figure}[H]
	\centering
	\includegraphics[scale=0.4]{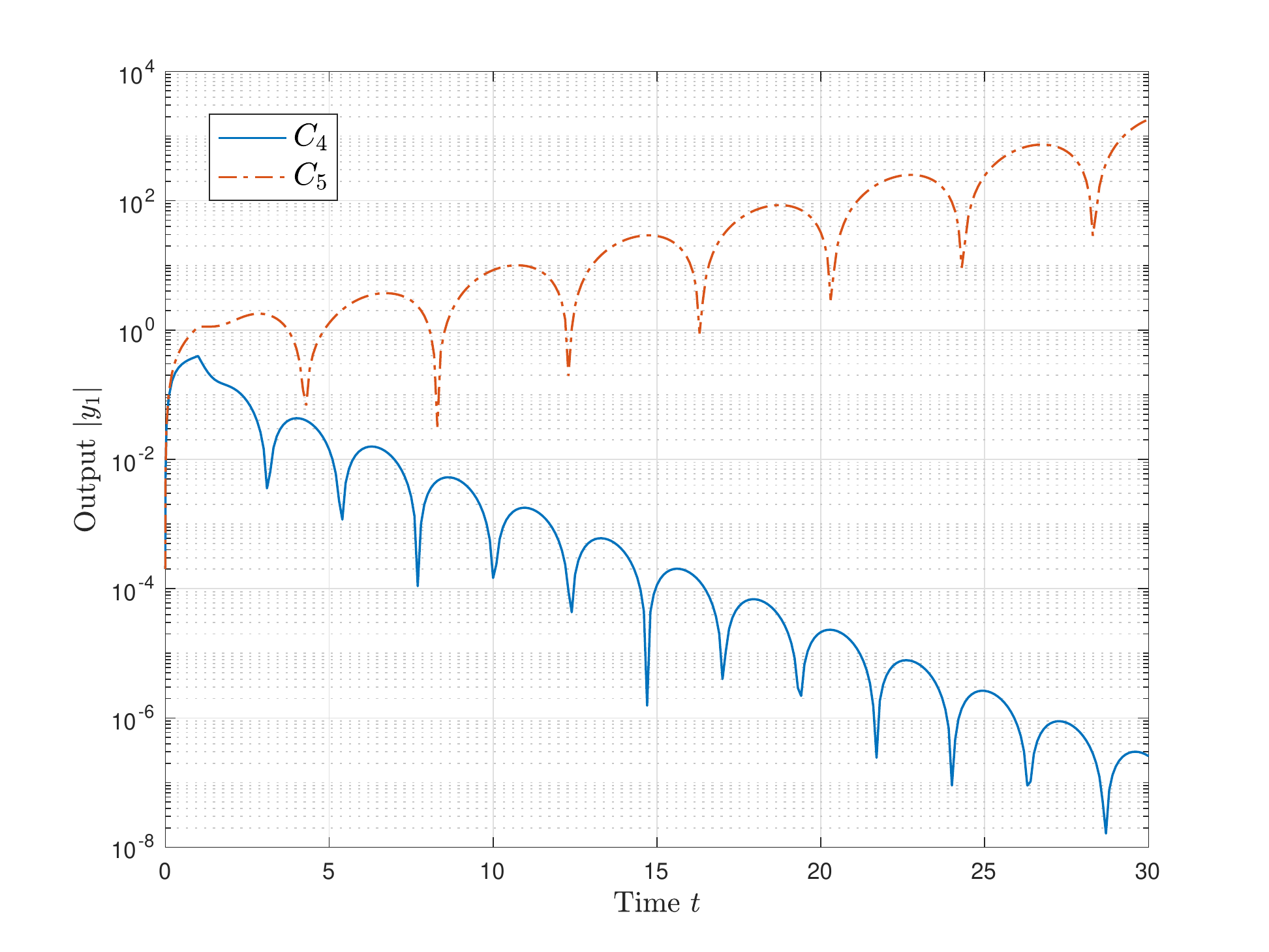}\\
	\caption{The amplitudes of output signals $y_1$ {(in terms of the log scale)} corresponding to different choices of $\bmC_i$, $i=4,5$, when $d_1$ is an impulse at time zero.}\label{figexampleNISNI}
\end{figure}
\black{Consider $\bmC_4=u\mapsto y\in\mathcal{N}_1$ with
$$\left\{\begin{array}{l}\dot{x}_1=x_2-u\\
	\dot{x}_2=-3x_1-\dfrac{x_2}{1+x_2^2}+u\\
	\dot{x}_3=-2x_3+4u\\
	y=x_2+x_3-4u\end{array}\right.
$$
and $\bmC_5=u\mapsto y\in\mathcal{N}_1$ with
$$\left\{\begin{array}{l}\dot{x}_1=x_2-u\\
	\dot{x}_2=-3x_1-\dfrac{x_2}{1+x_2^2}+u\\
	\dot{x}_3=-2x_3+u\\
	y=x_2+x_3-u\end{array}\right.,
$$
which can be equivalently represented as $\bmC_4=\bmP+4\bm{G}$ and $\bmC_5=\bmP+\bm{G}$, 
where $\bm{G}$ is LTI with transfer function $$\hat{G}=-\frac{s+1}{s+2}.$$
Since $\bmP$ is nonlinear NI and $\bm{G}$ is SNI, it follows by Proposition~\ref{prop:convex_NI}(b) that $\bmC_4$ and $\bmC_5$ are SNI. Furthermore, one can verify that $\bmC_4\in\mathcal{B}(\Xi_1,\epsilon)$ for each \black{$\epsilon\in(0,0.5]$} while there exists no $\epsilon>0$ such that $\bmC_5\in\mathcal{B}(\Xi_1,\epsilon)$. As shown in Fig.~\ref{figexampleNISNI}, $\nlcls_4$ produces an energy-bounded output signal given an impulse input signal, which is necessary for the feedback system to be stable. On the other hand, note that $\nlcls_5$ does not satisfy the premises in Theorem~\ref{thm:SNI2}, and the simulation also reveals that such a system is unstable.}

\section{Conclusions and Future Directions}\label{sec:conc}
This paper proposes an extension of negative imaginary systems theory to nonlinear systems framework. In contrast with the existing results based on state-space characterizations \cite{angeli2006CCW,Petersen2018CDC} or some special input-output index called ``phase'' \cite{chao2020nonlinear}, the proposed extension relies on a pure frequency-domain characterization of the NI property and develops a feedback stability condition for such systems via the theory of integral quadratic constraints. 

{Of future interest are generalizations to more general nonlinear systems (such as, unstable ones) and their feedback interconnections. In addition, as we have obtained that parallel interconnections preserve negative imaginariness, another possible future direction is to explore more useful properties of different interconnections between such systems.}

\bibliographystyle{apacite}  
\bibliography{Neg_Img}
\end{document}